\documentclass[aps,preprint]{revtex4}
\def\b{\begin{equation}}
\def\e{\end{equation}}

\begin{document}

\title
{Comments on ``The Euclidean gravitational action as black hole entropy, singularities, and spacetime voids''}

\author{Abhas Mitra}

\email {amitra@barc.gov.in}
\affiliation { Theoretical Astrophysics Section, Bhabha Atomic Research Centre, Mumbai, India}


\date{\today}






\date{\today}
\begin{abstract}
In a recent paper, Castro (J. Math. Phys., 49, 042501, 2008)\cite{1} has mentioned that in two papers, Mitra  (Found. Phys. Lett., 13, 543, 2002)\cite{2} and
( MNRAS, 369, 492, 2006)\cite{3}, (i)  attempted to show that a neutral point particle has zero gravitational mass, but (ii) his intended proof  
was faulty and can be ``bypassed''. In reality, none of these two papers offered any such proof and on the other hand this proof could be found
in (Adv. Sp. Res. 38(12), 2917, 2006)\cite{4, 5}  not considered/cited by Castro. This  shows that Castro critisized Mitra's ``proof'' without really going through it. We briefly revisit this ``proof'' to show that it is indeed correct and does not suffer from perceived shortcomings.   It is reminded that Arnowitt, Deser \& Misner's (ADM) previous work\cite{6} work suggested
that the gravitational mass of a {\em neutral point particle} is zero. It is pointed out that
that this result has important implications for Castro's work.
Further the  mass formula $M=\int 4\pi \rho r^2 dr$ used by Castro  may not be valid for the radial gauge he used, and the ``spacetime void'' inferred by Castro may be an artifact of the  discontinuos radial gauge used by him.  

\end{abstract}
\maketitle

\section{Introduction}
 The general general relativistic solution for the spacetime around a point mass may be given by\cite{7,8}
\b
ds^2=(1 - \alpha_0/ R) dt^2 
-(1-\alpha_0/ R)^{-1} dr^2 - 
 R(r)^2(d\theta^2 + \sin^2\theta d\phi^2)
\e
where $r$ is a general  radial coordinate. The variable $R(r)$ in the angular part is the circumference or area coordinate which means that the invariant area of a  two sphere
 around the center of symmetry is given by $4 \pi R^2$ independent of the precise form of $R(r)$.  This solution involves an integration constant $\alpha_0$ and comparison with Newtonian solutions suggest that $\alpha_0$ is related to the gravitational mass of the ``point mass'' in the following way:
\b
\alpha_0 = {2 GM_0\over c^2}
\e
If we take $G=c=1$, we will have $\alpha_0 = 2 M_0$. The subscript $0$ here denotes ``point mass'' vis-a-vis an {\em extendended} static spherical body of
gravitational mass $M$ and corresponding $\alpha = 2GM/ c^2$.

Choice of various forms of the function $R(r)$  corresponds to various choices of radial gauge and accordingly one can have various solutions. The simplest choice of the gauge here is
\b
r = R
\e
And this choice was due to Hilbert:
\b
ds^2=(1 - \alpha_0/ R) dT^2 
-(1-\alpha_0/ R)^{-1} dR^2 - 
 R^2(d\theta^2 + \sin^2\theta d\phi^2)
\e
though, in text books, the above solution is ascribed to Schwarschild. This Hilbert gauge is most natural and physical in the sense that the invariant surface area of a sphere around the point of symmetry is $4 \pi R^2$ rather than $4\pi r^2$. In this gauge, the ``point particle'', the source of gravity, is located at $R=0$
and its {\em surface area} is $A_p =0$. Note, the time interval, for this particular form of the metric has been designated by $dT$.

The Brillouin radial gauge, on the other hand, is given by\cite{7}
\b
R(r) = r + \alpha_0
\e
 while the actual Schwarzschild gauge is given by\cite{8}
\b
R(r)^3 = r^3 + \alpha_0^3; \qquad ~or~ r^3 = R^3 -\alpha_0^3
\e
In these latter cases, the point mass is believed to be sitting at the origin of the coordinate system, i.e., at $r=0$, and in such pictures, one has
\b
R(0) = \alpha_0
\e
so that, {\em the point mass has an invariant surface area} of 
\b
A_p = 4 \pi R(0)^2 = 4 \pi \alpha_0^2 >0~!
\e
if one would really assume that 
\b
\alpha_0 >0 
\e
So, for all such radial gauges, one would arrive at a rather grotesque picture where the {\em point mass has a finite surface are}!  And only for the Hilbert gauge,  one obtains the physical scenario where a point mass has zero surface area. Thus from such a view point, only the  Hilbert gauge (which however is ascribed to Schwarzschild in the literature)
leads to a physically meaningful solution for a ``point mass'' (atleast if one would indeed assume $\alpha_0 >0$). In fact Castro\cite{1} acknowledges this point: ``How is it {\em possible} for a point mass sitting at $r=0$ to have a non-zero area .. and a {\em zero} volume simultaneously?'' 

Hilbert's radial gauge however gives rise to a finite horizon area $A_H = 4 \pi \alpha_0^2$ if one would assume the {\em integration constant} $\alpha_0 >0$. However, Castro would like to consider a situation where both $A_H \to 0$, $A_p \to 0$. Thus he chose the following radial gauge which
seems to be some sort of modification of Brillouin's gauge:

\b
R(r) = r + \alpha_0 ~ \Theta(r); \qquad ~ r = R -\alpha_0 ~ \Theta(r)
\e
where $\Theta(r)$ is the Heaviside step function.

Castro also correctly points out that for a point particle scalar curvature must be a Dirac-$\delta$ function rather than 
${\cal R} =0$ as is mentioned in the literature.

It may however be reminded that the purpose of this ``comment'' is not to review or critisize Castro's work in general except for some passing comments. On the other hand, we would like
to focus attention on Castro's criticism of Mitra's work in  pp.11-12. Here Castro implied that, Mitra's purported proof that $\alpha_0 =0$ (for a neutral point mass) is incorrect. First it is pointed out that Mitra's proof to this effect is not at all presented in the papers cited\cite{2,3} by Castro. On the other hand, Castro did not cite the papers\cite{4,5} where such a proof was actually presented. First, this shows that Castro may not have at all gone through the said ``proof'' before dismissing it. For the benefit of the readers, it will be shown that this ``proof'' is correct and Castro's related criticism is misplaced. Further, it is emphasized that this proof removes much of the concern of Castro's work. It will be reminded  that previous work of  Arnowitt, Deser and Misner\cite{6} too suggests that a neutral point particle should have $M_0 =0$. Some basic error in Castro's work would be also pointed.

\section{Invariance of Proper 4-Volume}

For any curvilinear coordinate transformations, one has
\b
d\Omega = \sqrt{-g} ~d^4 x =Invariant 
\e
where $g = \det g_{ik}$
 For instance see. Eq.(2.5.6), pp. 36 of  Carmeli\cite{9} or pp. 99 of Weinberg\cite{10} or  pp. 223 of Landau-Lifshitz\cite{11}. This invariance has got nothing to do with choice of radial or any other gauge impllied by Castro, and on the other
hand, it is a property of tensor transformations.

If the determinant corresponging to the spatial section is $h$, one would have
\b
-g = h~ g_{00}
\e
so that
\b
\sqrt{-g} ~d^4 x = (\sqrt{-h}~ dx^1 ~dx^2 ~dx^3) ~(\sqrt{g_{00}} ~ dx^0) = d{\cal V} ~ d\tau
\e
where $d{\cal V}$ is the proper 3-volume element and $d\tau$ is the proper time element. Thus the invariant
\b
\sqrt{-g} ~d^4x = d\Omega = Proper ~Spacetime~Volume~Element
\e

 In fact it is valid in arbitrary dimentions too. However, if one is concerned with a situation which involves only spatial dimensions so that $g$ is positive then $-g$ of Eq.(11) should be replaced by $+g$. For an illustration see pp.170-171 of Hartle\cite{12}. 
Now by using this invariance of $d\Omega$, we revisit our proof that a neutral point particle has zero gravitational mass this simple straight forward proof below:

\subsection{The Proof Overlooked by Castro}
First consider the fact that for the diagonal vacuum Hilbert metric(4), one has
\b
g = -R^4 ~\sin^2\theta
\e
Now let us recall the Eddington-Finkelstein metric form
  of the vaccum Hilbert metric \cite{13,14}
\begin{eqnarray}
ds^2 =  \left(1 - {\alpha_0 \over R}\right)dT_*^2 \pm {2\alpha_0\over R} dT_* dR \\ \nonumber
- \left(1 +{\alpha_0\over R}\right)dR^2 - R^2(d\theta^2 + \sin^2\theta d\phi^2)
\end{eqnarray}
where
\b
T_*=T\mp \alpha_0 \log \left( {R\over \alpha_0} -1\right), ~R_*=R,~ \theta_* =\theta;~ \phi_* =\phi
\e
The corresponding metric coefficients are
\begin{eqnarray}
g_{{T_*} R_*}=-(1-\alpha_0/R), ~g_{R_* R_*}=(1+\alpha_0/R) \\ \nonumber
g_{{T_*} R_*}=g_{R_* T_*} 
=\alpha_0/R
\end{eqnarray}
In this case too, the determinant is same as $g$:
\b
g_* = -g_{\theta_* \theta_*} g_{\phi_* \phi_*}(g^2_{T_* R_*} -g_{T_* T_*}
 g_{R_* R_*}) = -R^4 \sin^2 \theta =g
\e
 Now let us apply the principle of {\em invariance of 4-volume} for the coordinate
systems ($t$, $R$, $\theta$, $\phi$) and ($t_*$, $R_*$, $\theta_*$, $\phi_*$): 
 \b
d\Omega = \sqrt{-g_*}~ dT_*~ dR_*~ d\theta_* ~d\phi_* = \sqrt{-g_*}~  dT~ dR~ d\theta~ d\phi
\e
Since $g_*= g, ~ R_* =R, ~ \theta_* =\theta, ~\phi_* =\phi$, we  obtain
\b
dT_* = dT
\e
 But from Eq.(17), we have,  
 \b
 dT_* = dT +  {\alpha_0\over R -\alpha_0}
 \e
 Eqs.(21) and (22) can be satisfied iff
\b
\alpha_0 =0 
\e
so that, for neutral ``point mass'' one has a zero gravitational mass: $M_0 =0$.

It is important to note that if instead of a ``point particle'' we would be considering a {\em finite}  static spherical body of gravitational mass
$M$, the above derivation would be invalid even though by Birchoff's theorem, the exterior spacetime would still be described by metric. This is so because, in such a case, exterior spacetime would be described only by metric(4)  and {\em not} by metric(16). This happens because the coordinate transformation(17)
is obtained by integrating the {\em vacuum} null geodesic all the way from $R=0$. And this is allowed only when the spacetime is indeed due to a point mass and not by a finite body. For a finite body, the {\em interior metric would be different} from what is indicated by Eq(4). and  hence metric(16)  which presumes the interior to be vacuum cannot be invoked.

Since, for a point mass $\alpha_0 =0$, the horizon area is zero and it contracts to the origin of the coordinate system without the application of any quantum mechanics or  quantum gravity.
Note that the original motivation of Castro too was see that horizon has zero area and lies at the origin of the coordinate system.

In the first line of pp.12, Castro\cite{1} writes that ``the measure $4\pi R^2 dR~dt$ is not invariant when we change the radial gauges...''

Again this statement is both confusing and incorrect. It must be borne in mind that if there is any change in the spatial coordinates (due to change of gauge or other reasons), the change actually takes place in ``spacetime'' description and hence the time coordinate too may undergo some transformation. For instance, let us consider metrics (1) and (4)  related through the coordinate transformations (5) or (6). It transpires that for all such cases, the metric determinant is same:
\b
g_H  =g_B= g_S = -R^4 \sin^2 \theta
\e
Then invariance of proper 4-volume element means
\b
dt ~dr~ d\theta~d \phi = dT ~ dR ~d\theta ~d\phi
\e
Thus the  Hilbert time element is related to other time elements in the following way:
\b
dt = {dR\over dr} ~dT
\e
For the Brillouin case, from Eq.(5), see that, $dR/dr =1$ and hence $dt =dT$. But, from Eq.(6), note that this  is not so in the Schwarzschild case:

\b
dt = {r^2\over R^2} ~ dT 
\e
\section{Inevitability of this result}
We know that the gravity at the horizon is supposed to be ``so strong that even lighy cannot escape it''. Let us first measure the strength or weakness of gravity by tKretschmann scalar, and for the given problem, one has
\b
K = {12 \alpha_0^2\over R^6}
\e
Note that the denominator here contains area coordinate $R$ rather than any general $r$ and $K =\infty$ only at $R=0$. This suggests that the genuine physical singularity lies at $R=0$ rather than at any general $r=0$. In turn, this tells that, the source of gravity is at $R=0$ rather than at arbitrary $r=0$. And this is again a strong argument that only the Hilbert gauge gives a physically meaningful picture. And of course, since $K$ is an invariant like $d\Omega$, it is independent of choice of coordinates and gauges.  

At the EH $R=\alpha_0$, one finds
\b
K^{EH} = {12\over  \alpha_0^4} 
\e

But now note that, if the integration constant $\alpha_0$ would be considered to be a free parameter, one can let $K$ to be arbitrarily small. For instance, for sufficiently large value of $\alpha_0$, one may have a situation where
\b
K^{EH} \ll K^{Earth}
\e
But if the weak gravity of Earth, Sun or Moon or even a Neutron star cannot trap light, how, a still waeker gravity would do so? Further, one can let
\b
K^{EH} \to 0 \qquad ~when~ \alpha_0\to \infty
\e
And this would mean that {\em light would get trapped in zero gravity}! Obviously, this would be unphysical, and consequently, $\alpha_0$ cannot be a free parameter. And as shown above, $\alpha_0 =0$ so that, actually
\b
K^{EH} = \infty
\e
and this is reason that even light gets trapped at the EH.
\subsection{Upper Limit on Proper Acceleration}
In his paper, Castro has correctly mentioned that quantum gravity should lead to an upper limit on proper acceletation:
\b
a_{Planck} = {c^2\over l_P}
\e
where $l_P$ is Planck length. But  in the spacetime around a  ``point particle'', the proper acceleration experienced by a test particle is
\cite{12}  (pp. 459) 
\b
a= \sqrt {-a^i a_i} = { G M_0\over R^2~\sqrt{1- \alpha_0/R}}
\e
where $a^i$ is the 4-acceleration. Again note that since $a$ is a scalar/invariant, this result is independent of coordinates and gauges. At the EH, $R=\alpha_0$, and hence, one would have
\b
a^{EH} = \infty
\e
Since $a$ is a physical scalar and is measurable, one expects that, it can blow up only at the genuine physical singularity  which lies at $R=0$ and where $K=\infty$. This, in turn, demands that coordinate radius of the horizon $\alpha_0 =0$ so that
the horizon and the central singularity are synonymous. Recall that we have already obtained this result $\alpha_0 =0$ from the invariance of $d\Omega$.

\subsection{Vanishing of Curvature Scalar}
By definition, for a point particle, the density and hence components of the energy momentum tensor $T_i^k$ would be singular at the location of the point particle. Since, $K^{EH}$ diverges at $R=0$ rather than at an arbitrary $r=0$, we expect a Dirac-$\delta$ singularity in the curvature scalar  at $R=0$\cite{15} 
\b
{\cal R}(R =0)= -(8 \pi G/c^4) T_i^i = - 4 GM_0\delta(R) /R^2 c^2 
\e
Depending on the model of the stress energy tensor one adopts, the numerical factors may differ in the foregoing expression\cite{16}. However, if one would calculate ${\cal R}$ using the Hilbert metric(4), one would obtain
\b
{\cal R} =0
\e
One necessary condition for the mutual consistentcy of Eqs. (36) and (37) is $M_0 =0$.

\section{Interior Solution}
If the point particle is question would be replaced by an extended distribution of mass energy, then, within this distribution, one must consider suitable {\em interior solution}. But the metrics (1) or (4) {\em do not represent} such interior solutions.  As far as static spherically symmetric interior solutions are concerned, all discussions are based on Hilbert radial gauge $r =R= cirumference ~ coordinate$. In literature, however, this is known as ``Schwarzschild Interior Solution''. It is only for this unique radial gauge, the gravitational mass of the body is given by
\b
M = \int 4 \pi \rho(R) ~ R^2 ~dR
\e
To the best of the knowledge, no corresponding study has ever been made for other choices of radial gauge. And in general, if $r \neq R$, there is no known formula for $M$. Thus, in general,
\b
M \neq \int 4 \pi \rho(r) ~r^2 ~dr
\e
if $r$ is an arbitrary radial coordinate. In his paper, Castro has replaced the ``point particle'' by a mass distribution extending upto $r=\infty$:
\b
\rho(r) = M {e^{-r^2/4 \sigma^2}\over (4 \pi \sigma^2)^{3/2}}
\e
and his $r \neq R$. Yet, he has used the mass formula appropriate only for Hilbert gauge $r=R$. 
\section{Spacetime Void?}
Castro\cite{1} considers a continuous classical spacetime metric. He does not explicitly use any quantum gravity where one might expect some spacetime irregularity on the scale of $l_P$. Yet Castro infers some ``spacetime void'' in the spacetime around a point particle. This might be due to the {\em discontinuos} radial gauge (10) used by Castro where
\b
{dR\over dr} = 1 + \alpha_0 \delta(r)
\e
and from Eq.(26), we see,
\b
dt_{Castro} = dT (1 + \alpha_0 \delta(r) )
\e
However, if the result $\alpha_0 =0$ would be taken into consideration, such strange notion of {\em spacetime void} would  get naturally eliminated.
\section{Conclusions}

$\bullet$ The mass formula used by Castro is inappropriate for any radial gauge where $r \neq R$.

$\bullet$ Invariance of proper 4-volume is a propery of tensor calculus and coordinate transformation. And this invariance indeed leads to $M_0 =0$
(for a neutral point particle). And when one recognizes this result, for a  point particle {\em all radial gauges become the same Hilbert gauge}.
Since the invariant surface area of two-spheres symmetric around the centre of symmetry, is $A = 4 \pi R^2$ (rather than $4 \pi r^2$), then it is most
matural that the source of gravity  is located at $R=0$ rather than at any arbitrary $r=0$. And this demands $M_0 =0$.

Long back, Arnowitt, Deser and Misner\cite{6} noted that
 
``  Thus as the interaction energy grows more negative, were a point reached where the total energy vanished, there could be no further interaction
energy, in contrast to the negative infinite self-energy of Newtonian theory. General relativity effectively replaces $m_0$ by $m$ in the interaction term: $m=m_0 - (1/2) G m^2/\epsilon$. Solving for $m$ yields $m = G^{-1} [-\epsilon + (\epsilon^2 + 2 G m_0 \epsilon)^{1/2}]$, { which shows that ${\mathbf m \to 0}$ as $\epsilon \to 0$}''.

``The correct result ${\mathbf m=0}$, thus indicates the lack of validity of the perturbation approach''\cite{1}.
Here $\epsilon$ may be viewed as the radius of a collapsing body  having a ``bare mass'' $m_0$ and a gravitational mass $m$ near the singularity. 

What these
authors mean is the following thing:

The fundamental source of ``mass energy'' is the fundamental interactions like electromagnetic, strong and weak. Gravitation is structure of spacetime
and is not a fundamental interaction in the above sense. However gravity couples to all mass energies and produces ``dressing'' in the form of
negative self-gravitational energy $E_g$. As the object tends to become a ``point particles'', $E_g \to unbounded$ and thus tends to nullify the entire
bare mass.  If the body would be neutral with nil electric, weak or strong ``charge'', the final gravitational mass of  the resultant ``point particle'' would be zero.

So, it is the infinite non-linearity of gravity and consequent growth of (negative) self-gravitational energies which may be leading to a net
 dressed (i.e., gravitational) massed energy which is nil.
 
 Thus the comment by Castro that Mitra's proof is incorrect and can be ``bypassed'' is invalid.
 
 Then the gauge used by Castro too becomes none other than Hilbert gauge. Then both the horizon and the point particle have zero surface area: $A =A_p =0$. And this is what Castro wanted to be. In such a case, the mass formula, used by him would become appropriate. And the {\em spacetime void} inferred by him in a purely classical theory based on continuous metric would vanish too.




\end{document}